# Risk Aversion in Non-Ergodic Systems

**Ihor Kendiukhov**
Faculty of Economics and Business Administration, Humboldt University of Berlin
Unter den Linden 6, 10099 Berlin
Germany
e-mail: kenduhov.ig@gmail.com

**Abstract** We show in a simulation when economic agents are subject to evolution (random change and selection based on the success in the estimation of the result of the gamble) they acquire risk aversive behavior. This behavior appears in the form of adjustment of their estimation of probabilities when calculating the expected value (ensemble average). It means that their subjective probabilities evolve in such a way that economic agents tend to assign lower probabilities to "good" events and higher probabilities to "bad" events. These subjective probabilities can be derived analytically by assuming that economic agents care about time average, not the ensemble average. Probabilities calculated based on this assumption are equal to the probabilities we get in evolutionary simulation. Furthermore, it appears that these subjective probabilities are equal to risk-neutral probabilities in mathematical finance. Hence, by taking into account that the environment in which economic agents operate is non-ergodic, we are able to calculate risk-neutral probabilities that are consistent with modern finance theory but are derived in a conceptually different way. It means that when we assume that agents try to predict time average, not ensemble average, the need for the concept of risk aversion disappears, since there is no distortion of subjective probabilities. Evolutionary simulations are quite a general method that can be applied for the determination of relevant measures of the outcome of a gamble under various conditions.

**JEL Classification Numbers** C61 · C63 · G11 · G12

**Declarations**

**Funding** This research did not receive any specific grant from funding agencies in the public, commercial, or not-for-profit sectors.
**Conflicts of interest/Competing interests** The authors declare that there is no conflict of interest.
**Availability of data and material** The authors confirm that the data supporting the findings of this study are available within the article and its supplementary materials.
**Code availability** The code is available in the appendix.

# 1 Introduction

In this paper, we follow the approach of Ole Peters and Murray Gell-Mann (Peters 2019; Peters, Gell-Mann 2016) to study economic agents` behavior under risk in the environments characterized by multiplicative dynamics. Namely, we focus on the study of time average as a relevant measure of the outcome of a gamble in the non-ergodic environment. It is shown that if expected value maximizers are subject to evolution (random mutation of subjective probabilities and selection based on the success in the prediction of the outcome of the gamble), they tend to become what is called risk aversive in modern economic literature. It means that their subjective probabilities evolve in such a way that economic agents tend to assign lower probabilities to "good" events and higher probabilities to "bad" events. It happens even if the probability of survival is 1 (the outcome of the multiplicative cannot become equal to 0). These subjective probabilities can be derived by assuming that economic agents do care about time average, not ensemble average, in such gambles (they put subjective probabilities when calculating ensemble average so that ensemble average is equal to time average under objective probabilities). Probabilities calculated based on this assumption are equal to the probabilities we get in evolutionary simulation. Furthermore, it appears that these subjective probabilities are equal to risk-neutral probabilities in mathematical finance (Hull 2009) if economic agents perceive time average as the risk-neutral discount rate. Although risk-neutral probabilities are derived in a very different way (based on no-arbitrage pricing), the formulas are surprisingly similar. There is a strong correspondence between risk-neutral probabilities of mathematical finance and subjective probabilities of non-ergodic environments.

Hence, ergodicity economics delivers risk-neutral probabilities which are consistent with modern finance theory but are derived in a conceptually different way. Q-probabilities of mathematical finance can be described as a special case of probability re-weighting under multiplicative dynamics. If we implement the approach of ergodicity economics (usage of time average), the need for probability re-weighting in models disappears.

However, there may be more sources of risk aversion, since it may be present even in ergodic environments.

The paper is structured as follows. Literature review is introduced in section 2. The model is built and analyzed in section 3. In subsection 3.1, the methodology and the model specifications are given. In subsection 3.2, the results of the simulations are described. In subsection 3.3, the connections between risk-neutral probabilities and obtained results are analyzed. Discussion is introduced in section 4, and conclusions are described in section 5. Simulations programming code is given in the appendix.

# 2 Literature review

Our analysis is based on a multiplicative gamble introduced by of Ole Peters and Murray Gell-Mann (Peters 2019; Peters, Gell-Mann 2016), which is an elementary demonstration of how ensemble averages (expected values) can be misleading indicators of probable outcomes of random processes. The game is indeed drosophila for financial theory, for it enables us to study a variety of real-world aspects of financial markets dynamics and investors behavior, and with its help, we are capable of deriving a theoretical understanding of many empirical laws, features, and pieces of evidence regarding the decision-making under uncertainty. However, the problem of the misuse of expected returns started being investigated even earlier (Hughson, Stutzer, Yung 2006).

Meder and colleagues (2019) show that in experiments people indeed tend to differentiate between ergodic and non-ergodic environments, and their behavior is risk aversive in non-ergodic environments. Obviously, it is an evolutionary adaptation emerged in a conceptually similar process of selection as we propose (survival of those who predict outcomes the best). Santos and colleagues (2009) review many important aspects of such an evolution of humans in the real world. It is shown also that some kind of risk aversion emerges in other species as well (Caraco, Martindale, Whittam 1980). It is likely that selection leading to risk aversion is quite a general phenomenon.

The concept of risk-neutral probabilities and no-arbitrage pricing is based on the works by Hull (2009, 2019), Hirsa, Neftci (2013), Shreve (2004), Rebentrost, Gupt, Bromley (2018), Horst (2018). Stochastic calculus developed by Itô (1944) is crucial for understanding dynamics in multiplicative gambles.

The notion of ergodicity, its definition and features are described by Lebowitz and Penrose (2008). The ergodicity economics framework is analyzed and presented here in comparison with classic expected utility theory by Neumann and Morgenstern (1947).

Probability re-weighting is a widespread approach in economics. We have already pointed out risk-neutral probabilities, but models in a similar spirit are widely applied in behavioral economics as well (Kahneman, Tversky 1979, 1992). Probability re-weighting, in general, is not limited to the modeling of risk aversion and is applied for the modeling of human biases.

Risk aversion itself is a very well-developed concept in modern economic theory. Its fundamentals are given by Arrow (1971).

The study of dynamic games cannot be separated from the study of time preferences and time discounting (Frederick, Loewenstein, O'Donoghue 2002). Risk aversion can be thus modeled either as the modification of probabilities ("risk-neutral probabilities") or as the modification of discount rate ("risk premium"). Discounting in the context of ergodicity economics is investigated by Adamou, Berman, Mavroyiannis, Peters (2019).

## 3 Results

### 3.1 Theory, methodology, and model

The ergodicity hypothesis can be written as (Birkhoff's equation):

$$\lim_{T \to \infty} \frac{1}{T} \int_0^T f(\omega(t))dt = \int_\Omega f(\omega)P(\omega)d\omega \qquad (1)$$

It tells that time average is equal to ensemble average. For detailed explanation of the formula, see Peters (2019).

If the system is ergodic, then the ensemble average (which is obviously equal to time average in this case) is a relevant measure of the expected return of the gamble. If not, it is more appropriate to use the time average.

The game used for the simulation is the following:

$$\Delta x = \begin{cases} \Delta x_H = +0.5x, & p_H = 0.5 \\ \Delta x_T = -0.4x, & p_T = 0.5 \end{cases} \qquad (2)$$

For detailed explanation of the gamble, again, see Peters (2019).

In this paper, we are going to show that in non-ergodic games, playing agents whose survival is conditioned on their success in predicting outcomes of the game, will naturally become what is called "risk aversive" in modern economic theory.

Firstly, let us denote expected value of the gamble as $\mu$. Then, we can rewrite $\Delta x_H$ and $\Delta x_T$ as $\mu + \sigma - 1$ and $\mu - \sigma - 1$ respectively. Thus, let us find time average of the game, which is also geometric mean:

$$E_t = (\mu - \sigma_1)^{1-p}(\mu + \sigma_2)^p \qquad (3)$$

where $\mu$ – ensemble average, $\sigma_1$ – deviation from ensemble average in case T, $\sigma_2$ – deviation from ensemble average in case H, $p$ – probability of H, $1 - p$ – probability of T. For this particular game, time average can be written in a shorter form:

$$E_t = \sqrt{\mu^2 - \sigma^2} \qquad (4)$$

where $\sigma$ is standard deviation of the outcome of the gamble.

It is already clear from formula (2) that if the agent uses time average as a criterion for decision making under risk, we will observe the pattern of risk aversion (negative impact of variation) if we define utility as a function of expected value $\mu$.

Now let us find such q-probabilities for ensemble average, that ensemble average is equal to time average calculated under objective probabilities:

$$E_e^q = E_t^p \qquad (5)$$

From now on, we shall call q-probabilities "risk-neutral probabilities". Firstly, it reflects the fact that risk aversion is incorporated into these probabilities. Secondly, as we will see below, it appears that these subjective probabilities are equal to risk-neutral probabilities in mathematical finance (Hull, 2009), if economic agents perceive time average as risk-neutral discount rate. There is a connection between risk-neutral probabilities of mathematical finance and subjective probabilities of non-ergodic environments.

If agent uses time average to predict the outcome of the game, risk-neutral probabilities will be:

$$(\mu - \sigma)(1 - q) + (\mu + \sigma)q = (\mu - \sigma_1)^{1-p}(\mu + \sigma_2)^p \tag{6}$$

$$q = \frac{(\mu - \sigma_1)^{1-p}(\mu + \sigma_2)^p - \mu + \sigma_1}{\sigma_1 \sigma_2} \tag{7}$$

For our game, $1 - q \approx 0{,}613, q \approx 0{,}387$.

When considering people making estimations of results of the gamble, we may find that they tend to be "irrationally" pessimistic about q-probabilities, but this "irrationality" vanishes when we discover how these probabilities are derived and constructed. Agents modify q-probabilities so, that they are able to use adequately the concept of ensemble average. Expected value works by multiplicative dynamics under "subjective" q-probabilities, not "real" p-probabilities.

The idea is to use natural selection to show this.

In this particular case, the evolution of 1000 breeding agents attempting to predict the outcome of the game was simulated. The process for a binomial gamble is the following:

1. 1000 agents are created. Their risk-neutral (q-probabilities) of $\Delta x_H$ and $\Delta x_T$ (see the description of the gamble) are equal to objective probabilities, namely 0,5 and 0,5 respectively.
2. Each agent is subject to "mutation", which changes his probability estimates randomly by adding to the previous probability estimates randomly generated values from the normal distribution with the expected value of 0.
3. 1000 steps of the gamble are simulated. The result of this simulation X is saved. The average rate of growth for one period is computed as $X^{\frac{1}{1000}}$ and saved as x.
4. Each agent takes ensemble average for the gamble based on his risk-neutral probabilities. Each ensemble average is compared to x. Those 10 agents whose differences between x and ensemble averages are the smallest, "survive". Other agents "die" (they are eliminated from further steps in the simulation). For 10 survived agents, average risk-neutral probabilities are computed, and one agent with these average risk-neutral probabilities is created (analog of breeding). The agent is now propagated to a thousand agents.
5. Steps 2,3,4 repeat 1000 times.

For the programming script and details of the simulations, please see the Appendix.

Why may some breeding analog be helpful? In principle, simulation of the selection would work without breeding as well. But breeding is an efficient invention of evolution which enabled organisms to transfer traits horizontally and to avoid the accumulation of harmful mutations. The presence of breeding makes evolution more efficient, and in our simulation, this is the case as well (Pontarotti et al. 2017). The benefits of breeding will be even more clear in the second simulation, where the value is multiplied by one of 100 randomly generated numbers each step of the gamble.

In other words, this simulation "kills" those who cannot predict outcomes of the gamble via ensemble average and "breeds" those who can do it given their q-probabilities. Such kind of simulation may be informative in different ways:

1. It enables us to understand whether risk aversion in non-ergodic environments can emerge naturally, without people understanding consciously the concept of ergodicity.
2. Successful natural selection simulated for a very large sample of agents shows that the ensemble average does not become a proper measure of the outcome under multiplicative dynamics even if the sample is large (in terms of real-life sizes).
3. It enables us to investigate the conditions under which risk-neutral agents can exist and act successfully in the markets.
4. It enables us to test various models attempting to predict successful strategies in dynamic gambles.

3.2 Simulations

Now let us turn to the simulation itself. In order to show the dynamics of the evolution, we plot *(1-q)/q* ratio by each period of evolution on the graph (figure 1). It always starts with 1, since initial "subjective probabilities" are equal to objective, which are 0,5 and 0,5.

The ratio *(1-q)/q* goes up and stabilizes then at the value of 1,59, which is ≈ 0.641/0,359. This is very close to the values we calculated for risk-neutral probabilities earlier.

The situation is similar in a more complicated game when we have 100 different numbers, each of which occurs with the probability of 1 percent. But now we compare the probability of the "worst" event W (the probability of the smallest number) with the probability of the "best" event B (the probability of the biggest number). p(W)/p(B) ratio grows, as predicted. Agents become more "pessimistic" (figure 2).

In both cases, we see that geometric mean is a relevant measure of the outcome of the gamble, and this result may seem trivial.

But geometric mean in this and similar cases is relevant as long as it delivers the value for time average, and not vice versa. Ergodicity economics framework delivers a criterion of when exactly geometric mean is a relevant measure of the outcome of the gamble and what else can be done in order to estimate the time average in non-ergodic processes.

Moreover, the geometric mean would be less relevant if we had more simultaneous gambles, though multiplicative as well. In general case, neither ensemble average nor time average is "the best" measure for gamble outcome, and in order to understand which one can be used to which extent, we need to investigate the process on ergodicity. Also, it is important to see whether natural selection itself can lead to such a pattern of risk aversion, which especially interesting in the case if people are able to distinguish intuitively ergodic and non-ergodic systems (Meder et al. 2019).

But risk aversive individuals become even far more risk aversive if we add the possibility of "death" to the game, in other words, if there is a chance that the wealth will be multiplied by 0 (figure 3). Of course, in this case, it is impossible to recover from zero because the dynamics are multiplicative. Hence, it is equivalent to death.

We see that the equilibrium q/(1-q) ratio increases by an order of magnitude in this case. Now we can switch to continuous probability distributions. In this case, risk aversion will arise as well.

For example, let us consider a very general case of normal distribution.

In the simulation, there is again some type of natural selection, but agents are equipped not with probabilities, but with parameters $\mu$ and $\sigma$ of normal distributions. These parameters are now subject to mutations.

Hence, on Y-axis there is another variable now, namely parameter $\mu$ of "subjective normal distribution" (figure 4).

Obviously, for each $\mu$ and $\sigma$ there is a subjective $\mu_s$ which is less than $\mu$. From stochastic calculus (Itô 1944), we know that

$$\mu_s = \mu - \frac{1}{2}\sigma \qquad (8)$$

Now we explain in more details why q-probabilities are "risk-neutral". They indeed not only "incorporate" risk in themselves but, also, they are deeply connected to the common definition of risk-neutral probabilities in finance theory (Hull 2009).

3.3 Results in the context of the concept of risk-neutral probabilities and risk aversion

From mathematical finance, we know that risk-neutral probability can be calculated this way:

$$q = \frac{e^r - d}{u - d} \qquad (9)$$

where $r$ – continuously compounded risk-free rate of return ($e^r$ can thus be considered as risk-free gross rate of return). Both $u$ and $d$ are defined such that:

$$X_1 = \begin{cases} uX_0, & p \\ dX_0, & 1-p \end{cases} \qquad (10)$$

Let $u > 1$, $d < 1$, and $u - (up + d(1-p)) = (up + d(1-p)) - d$ (symmetry). Then, we can rewrite formula (7) as:

$$q = \frac{(\mu-\sigma_1)^{1-p}(\mu+\sigma_2)^p - (\mu-\sigma_1)}{(\mu+\sigma_2)-(\mu-\sigma_1)} = \frac{(\mu-\sigma_1)^{1-p}(\mu+\sigma_2)^p - d}{u-d} \qquad (11)$$

It is surprisingly similar to the expression from finance theory for q-probabilities. There is a correspondence between both types of q-probabilities.

If formulas (9) and (11) are equivalent, then $(\mu - \sigma_1)^{1-p}(\mu + \sigma_2)^p$ should be equal to the risk-free gross rate of return $e^r$ which also is reasonable when agents use time average as a measure of the outcome of a gamble. Time average is equal to the risk-free gross rate of return. It reflects the fact that there is no risk premium if we consider time expected returns and not ensemble average returns.

People take care only of expected return, not risk, if the expected return is calculated as time average.

Or, if people take care only of the time average, then q-probabilities derived in this paper are equal to q-probabilities in mathematical finance because there should not be any risk premium in the rate of return $(\mu - \sigma_1)^{1-p}(\mu + \sigma_2)^p$.

Note that the formula for q-probabilities in finance is derived in a completely different manner, by using the no-arbitrage pricing approach, yet it delivers a very similar result. Hence, the approach described in the paper can be applied as a new method to derive q-probabilities necessary for multiple asset pricing models. By calculating the time average under p-probabilities and comparing it to ensemble average under q-probabilities, we are able to derive q-probabilities without the knowledge of risk-free rate and without the assumption of absence of arbitrage. Similarly, we can solve a reverse task: to find time average by given q-probabilities.

## 4 Discussion

In general, similar evolutionary simulations may be a useful tool for the resolution of arguments regarding the relevance of different measures of the outcome of a gamble under various conditions. Cases of multiple continuous statistical distributions can be tested as well. Cases of lognormal, Lévy and exponential distributions are of particular interest. Starting from toy examples of the binomial tree, we should gradually approach the conditions of the real world, namely the empirical reality of financial markets in our case.

Note that the "risk-neutralization" of distributions can be made in 2 ways: change of the expected value of the distribution (downward) and change of the form of the distribution (right-side asymmetry). We have slightly touched the first case, and the second case completely remains to be investigated.

A problem of a very different nature, yet crucial for every evolutionary modeling, is the selection of the type and the parameters of mutations. In our case, mutations are modeled by the normal distribution, but it may happen that the results of the simulation may be sensitive to mutations` parameters. For example, in the case of the normal distribution, big variance does not allow equilibrium to establish. Other problems may arise when modeling other types of mutations. The question is: how agents modify their beliefs in the real world?

Besides evolutionary simulation (random mutation combined with the selection), it may be interesting to conduct a "cognitive simulation" (updating beliefs regarding q-probabilities each round), for example, by using Bayes`s rule.

There is one more fundamental question regarding the role of expected utility theory in economics, which is also crucial for further development of our knowledge of decision making under uncertainty. Basically, expected utility theory can be adjusted as long as it needed to incorporate constantly emerging evidence regarding decision making under risk. But at least in some cases, there is a more natural and easy way to model economic agents` behavior without the attraction of additional entities such as probability re-weighting, risk aversion or utility functions concavity. Whereas expected utility theory in its extended versions may be equivalent in terms of predictive power to the models of ergodicity economics, multiplicative dynamics processes are described in many cases more simply by ergodicity economics.

## 5 Conclusions

Evolutionary simulations of agents attempting to predict the outcome of a gamble using expected value lead to the emergence of risk aversion in these agents. Namely, agents tend to assign higher values to the probabilities of "bad" events and lower values to the probabilities of "good" events. It turns out that these "subjective probabilities" are connected and under some reasonable conditions equal to risk-neutral q-probabilities derived by the no-arbitrage pricing approach. These q-probabilities can be also derived by the implementation of the ergodicity economics framework, assuming that agents use time average as a proper measure of the outcome of a gamble.

Evolutionary simulations are quite a general method that can be applied for the determination of relevant measures of the outcome of a gamble under various conditions.

An essential part of risk aversion can be explained by the fact that economic agents act in non-ergodic environments, yet some aspects of risk aversion may be explained by other reasons.

A very important question remains: what are the other sources of risk aversion? Whereas the usage of time averages gives an essential explanation of risk aversive behavior, we know (O'Donoghue, Somerville 2018) that risk aversion exists in ergodic environments as well, and it may be that some risk aversion remains even if we replace ensemble averages with time averages in non-ergodic environments. There may exist various factors rationalizing risk aversion of economic agents, such as for example, the possibility of "death" in a multiplicative gamble. It may be reasonable to attempt to determine the relative impact of each factor on risk aversion and then to see whether some risk aversion still remains if we take into account all factors. The presence of such residual risk aversion would mean potentially some cognitive bias. In other words: is there some irrational, "behavioral" risk aversion?

We should also take into account that in real-life scenarios, agents usually make a decision about portfolios of imperfectly correlated assets, not single assets. Although dynamics are multiplicative in many of these cases and the systems are non-ergodic, time average returns of a single asset may become less relevant measure under these conditions. In general, the more assets and the fewer periods we have, the more relevant is the ensemble average return as a measure of the outcome of the process. Thus, similar evolutionary simulations can be conducted for gambles consisting of many uncorrelated subgambles in order to investigate the impact of the simultaneous random multiplicative process on the optimal risk aversion.

Whereas the dynamics in financial markets is mostly multiplicative, there are important economic cases of additive dynamics as well. For example, wage dynamics may be additive. Evolutionary simulations may shed light on additive processes as well, especially if there are some critical lower bounds in the values additive process can take. There may be further attempts of the explanation of the regularities found by behavioral economics with the tools of ergodicity economics.

## References

Adamou, A., Berman, Y., Mavroyiannis, D., & Peters, O. (2019). Microfoundations of discounting. *Available at SSRN 3463229*.

Arrow, K. J. (1971). The theory of risk aversion. *Essays in the theory of risk-bearing*, 90-120.

Caraco, T., Martindale, S., & Whittam, T. S. (1980). An empirical demonstration of risk-sensitive foraging preferences. *Animal Behaviour*, *28*(3), 820-830.

Hughson, E., Stutzer, M., & Yung, C. (2006). The misuse of expected returns. *Financial Analysts Journal*, *62*(6), 88-96.

Frederick, S., Loewenstein, G., & O'donoghue, T. (2002). Time discounting and time preference: A critical review. *Journal of economic literature*, *40*(2), 351-401.

Hirsa, A., & Neftci, S. N. (2013). *An introduction to the mathematics of financial derivatives*. Academic press.

Horst, U. (2018). Introduction to Mathematical Finance.

Hull, J. (2009). Options, futures and other derivatives/John C. Hull. Upper Saddle River, NJ: Prentice Hall,.

Hull, J. (2019). Mathematical Finance: A Very Short Introduction: by Mark HA Davis, Oxford University Press (2019). Paperback. ISBN 978-0198787945.

Itô, K. (1944). Stochastic integral. Proceedings of the Imperial Academy, 20(8), 519-524.

Coolidge, J. L. (1925). *An introduction to mathematical probability* (pp. 61-76). Oxford.

Arrow, K. J. (1974). The use of unbounded utility functions in expected-utility maximization: Response. *The Quarterly Journal Of Economics*, *88*(1), 136-138.

Tversky, A., & Kahneman, D. (1979). Prospect theory: An analysis of decision under risk. *Econometrica*, *47*(2), 263-291.

Lebowitz, J. L. & Penrose, O. (2008). Modern ergodic theory. *Physics Today 26*, 23–29

Lee, M. D., & Wagenmakers, E. J. (2013). Bayesian Cognitive Modeling: A Practical Course 1–129.

Buchanan, M. (2013). Gamble with time. *Nature Physics*, *9*(1), 3-3.

Meder, D., Rabe, F., Morville, T., Madsen, K. H., Koudahl, M. T., Dolan, R. J., ... & Hulme, O. J. (2019). Ergodicity-breaking reveals time optimal economic behavior in humans. *arXiv preprint arXiv:1906.04652*.

Von Neumann, J., & Morgenstern, O. (1947). Theory of games and economic behavior, 2nd rev.

Peters, O. (2011). The time resolution of the St Petersburg paradox. *Philosophical Transactions of the Royal Society A: Mathematical, Physical and Engineering Sciences*, *369*(1956), 4913-4931.

O'Donoghue, T., & Somerville, J. (2018). Modeling risk aversion in economics. Journal of Economic Perspectives, 32(2), 91-114.

Peters, O. & Adamou, A. (2018). Lecture notes. *Ergodicity Economics* https://ergodicityeconomics.com/lecture-notes/ .

Peters, O., & Adamou, A. (2011). Leverage efficiency. *arXiv preprint arXiv:1101.4548*.

Peters, O. & Adamou, A. (2018). The time interpretation of expected utility theory. Preprint at https://arxiv.org/abs/1801.03680.

Peters, O., & Gell-Mann, M. (2016). Evaluating gambles using dynamics. *Chaos: An Interdisciplinary Journal of Nonlinear Science*, *26*(2), 023103.

Peters, O. (2019). The ergodicity problem in economics. *Nature Physics*, *15*(12), 1216-1221.

Pontarotti, P. (Ed.). (2017). Evolutionary biology: self/nonself evolution, species and complex traits evolution, methods and concepts. Springer.

Rebentrost, P., Gupt, B., & Bromley, T. R. (2018). Quantum computational finance: Monte Carlo pricing of financial derivatives. Physical Review A, 98(2), 022321.

Santos, L. R., & Chen, M. K. (2009). The evolution of rational and irrational economic behavior: evidence and insight from a non-human primate species. In *Neuroeconomics* (pp. 81-93). Academic Press.

Shreve, S. E. (2004). Stochastic calculus for finance II: Continuous-time models (Vol. 11). Springer Science & Business Media.

Tversky, A., & Kahneman, D. (1992). Advances in prospect theory: Cumulative representation of uncertainty. *Journal of Risk and uncertainty*, *5*(4), 297-323.

**Appendix**

Evolutionary algorithm simulation scripts. All scripts a written in Python 3.

*Algorithm 1. Binomial gamble*

```python
import random
import random as wr
import numpy as np
import matplotlib.pyplot as plt
n=1000 #number of rounds of the evolution
p=1 #number of the simulation of x
d=1000 #iterations of x multiplications
m=100 #number of agents
y=2
v1=0.01
values=[0.6, 1.5]
average=0
deviation=0.001
prob=0.5
subjective_p=[prob, 1-prob]
weights=[prob, 1-prob]
sp1=0.01
number=1
ratios=[]
ratio0=1
for i in range(0,n):
    x=1
    players=[]
    differences=[]
    for k in range(0,m):
        player=[]
        for i1 in range (0,y):
            r=wr.gauss(average, deviation)
            if r+subjective_p[i1]>0:
```

```
                subjective_p[i1]=subjective_p[i1]+r
            else:
                subjective_p[i1]=0
        for i1 in range (0,y):
            subjective_p[i1]=subjective_p[i1]/sum(subjective_p)
            player.append(subjective_p[i1])
        u=0
        for i1 in range (0,y):
            u=subjective_p[i1]*values[i1]+u
        player.append(u)
        players.append(player)
    list=[]
    for j in range(0,p):
        x=1
        for i in range(0,d):
            a=random.choices(values, weights, k=1)
            a=a[0]
            x=x*a
        list.append(x)
    x=sum(list)/len(list)
    x=x**(1/d)
    parameters1=players[0]
    utility1=parameters1[y]
    for l in range(0,m):
        parameters=players[l]
        utility=parameters[y]
        difference=abs(utility-x)
        #print(u, x)
        differences.append(difference)
    differences_copy=differences
    indices=[]
    new_players=[]
    for v in range(0,10):
        winner=min(differences_copy)
        ind=differences.index(winner)
        differences_copy.remove(winner)
        new_player=players[ind]
        new_players.append(new_player)
        indices.append(ind)
    new_players=np.array(new_players)
    for i3 in range (0,y):
        subjective_p[i3]=sum(new_players[:, i3].tolist())/len(new_players[:, i3].tolist())
```

```
    try:
        print(subjective_p[0]/subjective_p[y-1])
        ratio1=subjective_p[0]/subjective_p[y-1]
    except:
        print('infinity')
        ratio1=ratio0
    ratios.append(ratio1)
    ratio0=ratio1
a=(subjective_p[0])
b=(subjective_p[y-1])
print(a,b)
ratio=(subjective_p[0])/(subjective_p[y-1])
plt.plot(ratios, 'r')
plt.show()
print(utility1-x)
print(ratio)
```

*Algorithm 2. The gamble with 100 possible outcomes.*

```
import random
import random as wr
import numpy as np
import matplotlib.pyplot as plt
n=1000
m=100
p=1
d=1000
y=100
v1=0.01
values=[]
average=0
deviation=0.00001
subjective_p=[]
sp1=0.01
number=1
ratios=[]
ratio0=1
for i2 in range (0,y):
    subjective_p.append(sp1)
for z in range (0,y):
    values.append(v1)
    v1=v1+0.02
for i in range(0,n):
    x=1
```

```python
players=[]
differences=[]
for k in range(0,m):
    player=[]
    for i1 in range (0,y):
        r=wr.gauss(average, deviation)
        if r+subjective_p[i1]>0:
            subjective_p[i1]=subjective_p[i1]+r
        else:
            subjective_p[i1]=0
    for i1 in range (0,y):
        subjective_p[i1]=subjective_p[i1]/sum(subjective_p)
        player.append(subjective_p[i1])
    u=0
    for i1 in range (0,y):
        u=subjective_p[i1]*values[i1]+u
    player.append(u)
    players.append(player)
list=[]
for j in range(0,p):
    x=1
    for i in range(0,d):
        a=random.choice(values)
        x=x*a
    list.append(x)
x=sum(list)/len(list)
x=x**(1/d)
parameters1=players[0]
utility1=parameters1[y]
for l in range(0,m):
    parameters=players[l]
    utility=parameters[y]
    difference=abs(utility-x)
    differences.append(difference)
differences_copy=differences
indices=[]
new_players=[]
for v in range(0,10):
    winner=min(differences_copy)
    ind=differences.index(winner)
    differences_copy.remove(winner)
    new_player=players[ind]
    new_players.append(new_player)
```

```
            indices.append(ind)
        new_players=np.array(new_players)
        for i3 in range (0,y):
            subjective_p[i3]=sum(new_players[:, i3].tolist())/len(new_players[:, i3].tolist())
        try:
            print(subjective_p[0]/subjective_p[y-1])
            ratio1=subjective_p[0]/subjective_p[y-1]
        except:
            print('infinity')
            ratio1=ratio0
        ratios.append(ratio1)
        ratio0=ratio1
    print(subjective_p)
    ratio=(subjective_p[0])/(subjective_p[y-1])
    plt.plot(ratios, 'r')
    plt.yscale('log')
    plt.show()
    print(utility1-x)
    print(ratio)
```

*Algorithm 3. The gamble with 100 possible outcomes and possibility of "death".*
```
import random
import random as wr
import numpy as np
import matplotlib.pyplot as plt
n=3000
m=100
p=1
d=1000
y=100
v1=0
values=[]
average=0
deviation=0.00001
subjective_p=[]
sp1=0.01
number=1
ratios=[]
ratio0=1
for i2 in range (0,y):
    subjective_p.append(sp1)
for z in range (0,y):
```

```
      values.append(v1)
      v1=v1+0.02
   for i in range(0,n):
      x=1
      players=[]
      differences=[]
      for k in range(0,m):
         player=[]
         for i1 in range (0,y):
            r=wr.gauss(average, deviation)
            if r+subjective_p[i1]>0:
               subjective_p[i1]=subjective_p[i1]+r
            else:
               subjective_p[i1]=0
            player.append(subjective_p[i1])
         u=0
         for i1 in range (0,y):
            u=subjective_p[i1]*values[i1]+u
         player.append(u)
         players.append(player)
      list=[]
      for j in range(0,p):
         x=1
         for i in range(0,d):
            a=random.choice(values)
            x=x*a
         list.append(x)
      x=sum(list)/len(list)
      x=x**(1/d)
      parameters1=players[0]
      utility1=parameters1[y]
      for l in range(0,m):
         parameters=players[l]
         utility=parameters[y]
         difference=abs(utility-x)
         differences.append(difference)
      differences_copy=differences
      indices=[]
      new_players=[]
      for v in range(0,10):
         winner=min(differences_copy)
         ind=differences.index(winner)
         differences_copy.remove(winner)
```

```
            new_player=players[ind]
            new_players.append(new_player)
            indices.append(ind)
         new_players=np.array(new_players)
         for i3 in range (0,y):
            subjective_p[i3]=sum(new_players[:, i3].tolist())/len(new_players[:, i3].tolist())
         try:
            print(subjective_p[0]/subjective_p[y-1])
            ratio1=subjective_p[0]/subjective_p[y-1]
         except:
            print('infinity')
            ratio1=ratio0
         ratios.append(ratio1)
         ratio0=ratio1
      print(subjective_p)
      ratio=(subjective_p[0])/(subjective_p[y-1])
      plt.plot(ratios, 'r')
      plt.yscale('log')
      plt.show()
      print(utility1-x)
      print(ratio)
```

*Algorithm 4. The gamble with normal distribution.*

```
import random
import random as wr
import numpy as np
import matplotlib.pyplot as plt
n=3000
m=100
p=1
q=1000
d=1000
y=2
v1=0
values=[]
average=0
deviation=0.000001
average1=1.1
deviation1=0.1
subjective_p=[average1,deviation1]
sp1=0.01
number=1
```

```
av=[]
ratio0=1
for i2 in range (0,y):
   subjective_p.append(sp1)
for z in range (0,y):
   values.append(v1)
   v1=v1+0.02
for i in range(0,n):
   x=1
   players=[]
   differences=[]
   for k in range(0,m):
      player=[]
      for i1 in range (0,y):
         r=wr.gauss(average, deviation)
         subjective_p[i1]=subjective_p[i1]+r
         player.append(subjective_p[i1])
      u=subjective_p[0]
      player.append(u)
      players.append(player)
   list=[]
   for j in range(0,p):
      x=1
      for i in range(0,d):
         a=wr.gauss(average1, deviation1)
         x=x*a
      list.append(x)
   x=sum(list)/len(list)
   x=x**(1/d)
   parameters1=players[0]
   utility1=parameters1[y]
   for l in range(0,m):
      parameters=players[l]
      utility=parameters[y]
      difference=abs(utility-x)
      differences.append(difference)
   differences_copy=differences
   indices=[]
   new_players=[]
   for v in range(0,10):
      winner=min(differences_copy)
      ind=differences.index(winner)
      differences_copy.remove(winner)
```

```
            new_player=players[ind]
            new_players.append(new_player)
            indices.append(ind)
        new_players=np.array(new_players)
        for i3 in range (0,y):
            subjective_p[i3]=sum(new_players[:, i3].tolist())/len(new_players[:, i3].tolist())
        print(subjective_p[0])
        av.append(subjective_p[0])
    print(subjective_p)
    ratio=(subjective_p[0])/(subjective_p[y-1])
    plt.plot(av, 'r')
    plt.show()
    print(utility1-x)
    print(ratio)
```

**Figures**

**Fig. 1** (1-q)/q ratio versus period of evolution

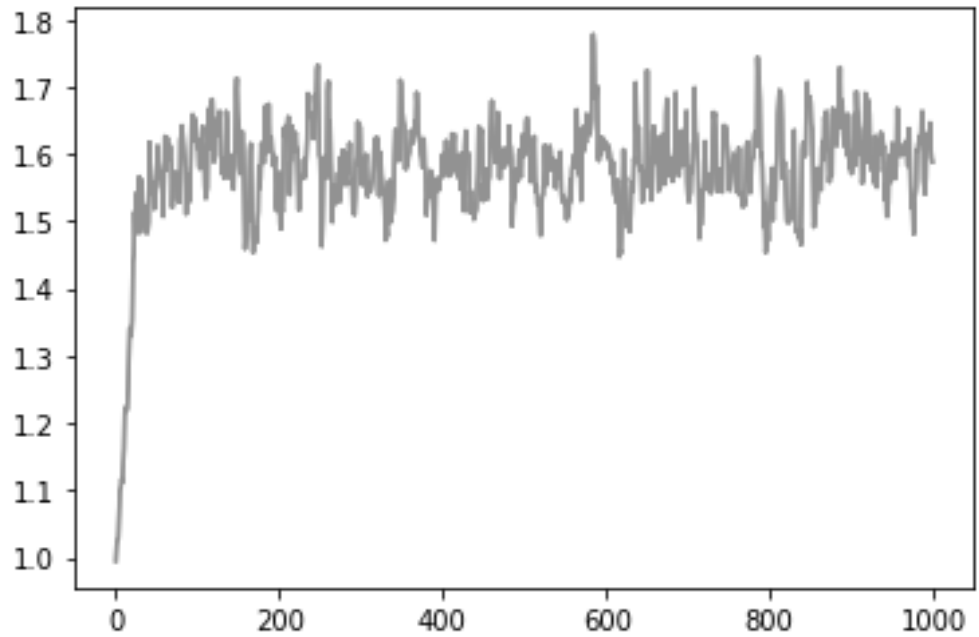

**Source**: Own results. Appendix. Algorithm 1. Created via Matplotlib

**Fig. 2** (1-q)/q ratio versus period of evolution

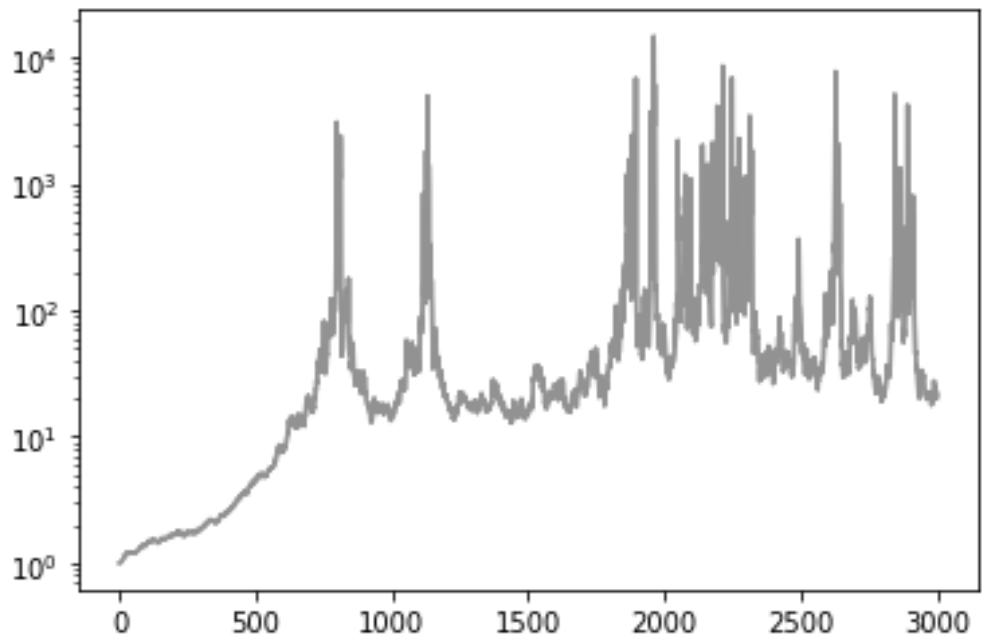

**Source:** Own results. Appendix. Algorithm 2. Created via Matplotlib

**Fig. 3** (1-q)/q ratio versus period of evolution

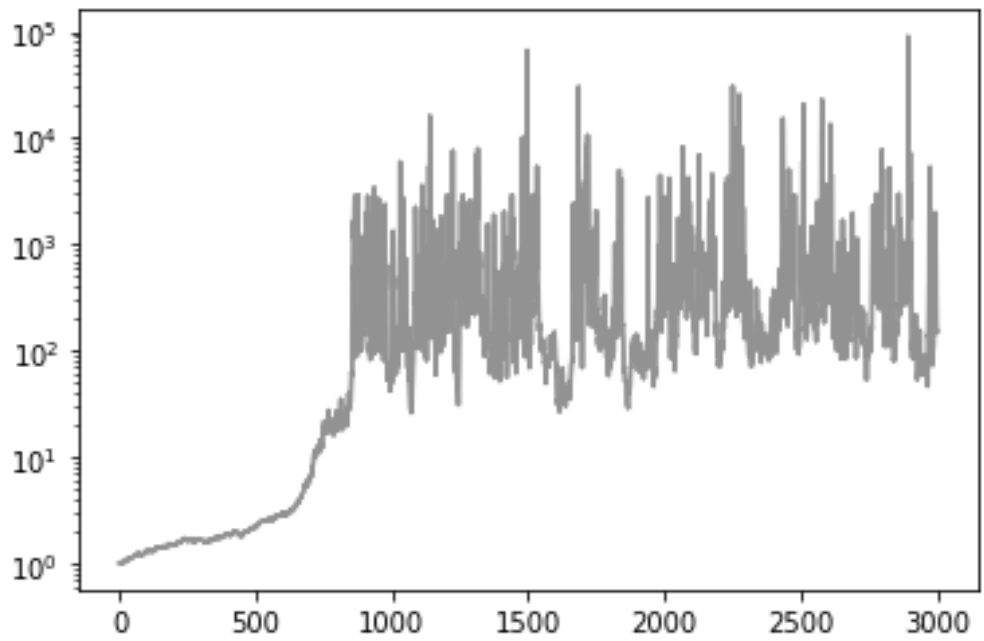

**Source:** Own results. Appendix. Algorithm 3. Created via Matplotlib

**Fig. 4** Parameter μ versus period of evolution

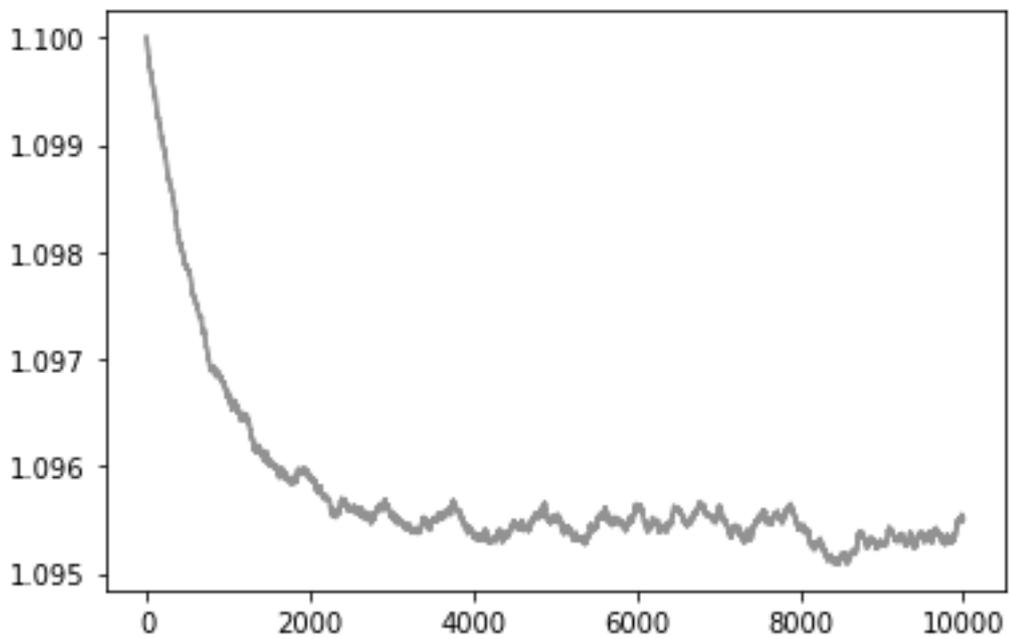

**Source:** Own results. Appendix. Algorithm 4. Created via Matplotlib